\documentclass[12pt,a4paper]{article}
\usepackage{amssymb,amsfonts,amsmath}
\usepackage{color}
\usepackage{epsfig, euscript, graphics}

\begin{document}

\title{Quantum versus classical entanglement: eliminating the issue of quantum nonlocality}

\author{Andrei Khrennikov\\ 
Linnaeus University, International Center for Mathematical Modeling\\  in Physics and Cognitive Sciences
 V\"axj\"o, SE-351 95, Sweden}

\maketitle

\abstract{We analyze interrelation of quantum and classical entanglement. The latter notion is widely used in classical optic simulation of some quantum-like features of light. We criticize the common interpretation that ``quantum nonlocality'' is the basic factor  
differing quantum and classical realizations of entanglement. Instead, we point to the breakthrough Grangier et al.  experiment on coincidence detection which was done in 1986 and played the crucial role in rejection of (semi-)classical field models in favor of quantum mechanics. Classical entanglement sources produce light beams with the coefficient of second order coherence $g^{(2)}(0) \geq 1.$ This feature of classical entanglement is obscured by using intensities of signals in different channels, instead of counting clicks of photo-detectors. Interplay between intensity and clicks counting is not just a technicality. We elevate this issue to the high foundational level.}

\section{Introduction}

The classical electromagnetic field can be successfully used to model some basic features of genuine quantum physical systems (see, e.g., \cite{NL0,NL,NLG} and recent review \cite{REV} and the references herein). In particular, classical field modeling is a helpful tool for  simulation in quantum information theory. However, 
the foundational output of quantum-like modeling with classical light and other types of waves is not straightforward. In this note, I would like to discuss the foundational meaning of so-called ``classical entanglement'' and the widely spread view that ``quantum nonlocality'' 
(whatever it means) plays the crucial role in distinguishing classical and genuine quantum entanglements. The ``quantum nonlocality'' viewpoint was clearly formulated by Spreeuw in  widely  cited paper \cite{NL}  which is often mentioned by experimenters for foundational justification of their activity:

\medskip

\footnotesize{``It is found that the model system ({\it classical electromagnetic 
field}) can
successfully simulate most features of entanglement, but fails
to simulate quantum nonlocality. Investigations of how far
the classical simulation can be pushed show that quantum
nonlocality is the essential ingredient of a quantum computer,
even more so than entanglement. The well known problem
of exponential resources required for a classical simulation of
a quantum computer, is also linked to the nonlocal nature
of entanglement, rather than to the nonfactorizability of the
state vector.''}

and then

\footnotesize{``However, the ({\it classical-quantum}) analogy  fails to produce effects
of quantum nonlocality, thus signaling a profound
difference between two types of entanglement: (i) ``true,''
multiparticle entanglement and (ii) a weaker form of entanglement
between different degrees of freedom of a single
particle. Although these two types look deceptively
similar in many respects, only type (i) can yield nonlocal
correlations. Only the type (ii) entanglement has a
classical analogy.''}

[Comments in italic were added by the author of the present paper.]

However, one can proceed without referring to mysterious quantum nonlocality - by taking into 
account that quantum theory is about acts of observations. These acts are characterized 
by {\it individuality and discreteness.}  This crucial point in understanding of  quantum theory was missed by authors discussing classical 
entanglement.  They missed that the main deviation of classical light models from quantum theory is not only in the states, but in 
descriptions of measurement procedures.\footnote{As a comment on the first version \cite{ARV} of this paper, I received email from Gerd Leuchs with the reference to recent review \cite{GL}. Authors of this review do not refer to quantum nonlocality to distinguish classical and ``true quantum entanglement''. Their position is closer to my own. We shall discuss it in more detail in section \ref{CS}. } The classical and semiclassical descriptions of measurements are based on intensities of 
signals in different channels. The quantum description of measurements is based on counting the discrete events, clicks of detectors,
with the aid of the Born's rule. Operating with intensities obscures  the problem of  {\it coincidence detection.}
We recall that  quantum theory predicts that the relative probability of coincidence detection given by the coefficient of second order coherence $g^{(2)}(0)$ is zero (for one photon states), but in (semi-)classical models  $g^{(2)}(0) \geq 1.$ 

\medskip

{\it Genuine quantum theory differs from classical light models reproducing quantum correlations 
not by ``quantum nonlocality'', but by the magnitude of  second order coherence.} Classical and semiclassical models were 
rejected long ago as the result of Grangier  et al.  \cite{GR} experiment on coincidence detection (see \cite{GRT} for 
historical review on such experiments).

\medskip

In fact, the main issue is the difference between classical and quantum superpositions, not entanglements. Our message is that each state-superposition has to be endowed with a proper measurement procedure. Otherwise superposition is just ``thing-in-itself''. Superposition endowed  with the classical measurement procedure crucially differs from superposition endowed with the quantum measurement procedure. This difference is elevated to the level of entangled states, classical vs. quantum, without measurement procedures they are also just   ``things-in-themselves'' and as such they are not interesting for physics.

Finalizing the introduction, we stress that``quantum nonlocality'' is really misleading notion. As  shown in \cite{BELL}, the Bell tests can be consistently interpreted in the purely quantum theoretical framework (without any coupling to Bell's hidden variables theory 
\cite{B1,B2,B3,CHSH,Jaeger1}) as statistical tests of local incompatibility of quantum observables, i.e., as tests of the most fundamental principle of quantum mechanics, {\it the complementarity principle} 
\cite{BR0} (see also \cite{PL1,PL2,Jaeger1}). For reader's convenience, the compact presentation of the ``Bohr against Bell argument'' is given 
in section \ref{NLB}.

\section{Nonlocality mess}

Nowadays playing with the notion ``quantum nonlocality''  is the real mess. 
People widely operate with this notion  and often without any specification
on its meaning. We briefly recall the history of its appearance.

The starting point of propagating of quantum nonlocality through the quantum community 
was the EPR-paper \cite{EPR}.  Here nonlocality in the form of a spooky action at a distance was counted as a possible 
alternative to incompleteness of quantum mechanics. The EPR-argument was based on invention of elements of reality and counterfactual reasoning. This reasoning can lead to the idea on mystical a spooky  action at a distance. Bohr replied Einstein \cite{BR} by pointing that 
EPR's criterion of physical reality becomes ambiguous in quantum physics (see \cite{SPD} for details of this debate).

Einstein and Bohr did not understand each other, because they behaved towards quantum mechanics in the totally different ways. For Bohr, quantum mechanics is observational theory, it is about measurements performed by classical measurement apparatuses on microsystems. In modern terminology, quantum mechanics is an epistemic theory \cite{HARALD}; it is about extraction of knowledge
about nature with the aid of measurements. For Einstein, quantum mechanics (as any physical theory) was a descriptive theory providing consistent and complete description of nature. Philosophers also use the notion of ontic theory, i.e., theory describing nature as it is 
- when nobody looks at it (see also \cite{KHERTZ}). 

For me, the root of disagreement between Einstein and Bohr can be found already in the interpretations of measurement on a single system.
(Consideration of compound systems and the EPR-states had just strengthen  this disagreement.) For Bohr, quantum mechanics generates predictions on outputs of interaction of a quantum system and a measurement apparatus; for Einstein, quantum mechanics (as any ``good physical theory'') should generate prediction of ``real physical properties of a system'' (see also section \ref{RWR}).

In any event, Einstein's message on a spooky  action at a distance approached and excited the quantum community. And at the same time, the seeding issue of (in)completeness of quantum mechanics was totally forgotten. (Only philosophers continue to debate the EPR-paper \cite{EPR}
from the  completeness-incompleteness viewpoint, see, e.g., \cite{SPD} for the most recent analysis of this issue).

By criticizing the interpretational output of extended research on classical entanglement, I only criticize coupling to  mystical quantum nonlocality. As  shown in  \cite{BELL} (see also \cite{Accardi}-\cite{Griffiths} and section \ref{NLB}), quantum mechanics by itself has no coupling to such kind of nonlocality. (This statement is also strongly supported by quantum field theory, e.g., 
\cite{BS,Haag}.) At the same time, a subquatum  theory can in principle be nonlocal, as Bohmian mechanics and other theories with hidden variables considered by Bell \cite{B1,B2,B3}. However, generally, in spite of the Bell theorem, a subquantum theory can be free of nonlocality of a spooky  action at a distance type; see \cite{PCT1,PQ} and Appendix 1 for {\it prequantum classical statistical field theory} (PCSFT). The latter  is the classical random field model beyond quantum theory. (Coupling between PCSFT and quantum mechanics is not
so straightforward  as in the Bell framework \cite{B1,B2,B3}). PCSFT pretends \cite{PCT1,PQ} that genuine quantum systems can be mathematically represented by classical random fields. So, it is not a part the classical entanglement project. It is 
a part of extensive studies on classical probabilistic reconstruction of quantum theory (see, e.g., \cite{Feynman}-\cite{KC7}). 

Thus I also contributed to random field modeling of quantum correlations. Therefore generally I am sympathetic to the classical entanglement project. Moreover, by reading Spreeuw's paper \cite{NL}, I had the impression that, in fact, by writing about ``quantum nonlocality'' he had in mind the correlations of spatially separated signals, as,e.g., in radiophysics (cf. with the above discussion on quantum nonlocality mess).

\section{Inter-intra system  versus quantum-classical entanglements}

As stated in recent review \cite{REV}, \footnotesize{``...the name classical entanglement denotes the occurrence of some mathematical and physical aspects of quantum entanglement in classical beams of light.  ... the term ‘classical’ in the name classical entanglement, indicates the nonquantum nature of the excitation of the electromagnetic field.  .... A typical example thereof is given by a collimated optical beam with non-uniform polarization pattern.''} We continue by citing \cite{NL}:
\footnotesize{``It should be noted that the choice of optical waves is not essential for the analogy. Other classical
waves such as sound, water waves, or even coupled pendula could be used in principle.''}

In short, classical entanglement is associated with the ``nonquantum nature of the excitation of the electromagnetic field'' or  
modes of classical waves of any origin. Generally, classical systems of any origin can be considered, see, e.g., \cite{AL} on entanglement 
of classical Brownian motions. 

This paper is directed against the statement that the difference between classical  and genuine quantum entanglements is due to quantum nonlocality. 

Now we remark that comparison classical-quantum entanglements
is typically coupled to comparison intra-inter system entanglements. 
Intra-entanglement is between degrees of freedom of a single system and  inter-entanglement is between degrees of freedom of two 
systems, $S_1$ and $S_2.$ 

Classical entanglement modeling 
is possible only in the intrasystem context  \cite{NL0,NL,NLG,REV}.  One may conclude that this feature of 
 entanglement plays the crucial role in distinguishing classical and  quantum entanglements.
This reasoning also leads to conclusion that only intersystem entanglement is ``true quantum entanglement''
(since intrasystem entanglement can be generated even with classical fields).  

In this paper, we demonstrate that classical intrasystem entanglement differs fundamentally not only from quantum intersystem entanglement, but even from quantum intrasystem entanglement. Thus, comparison classical-quantum entanglements has no relation to intra-inter comparison
(and, hence, no relation to  quantum nonlocality).    

This comparative analysis of inter-intra system  versus quantum-classical entanglements can be completed by the following remark.
The impossibility of classical representation of intersystem entanglement is related only to the very spacial class of 
the field models elaborated in the classical entanglement project \cite{REV} (cf. \cite{PCT1,PQ}: in PCSFT, both types of entanglement 
(intra and inter) can be realized, but they have different mathematical representations,  see Appendix 1  for further discussion). 

Finally, we note that the intra-inter system difference of entanglements  is invisible in the quantum theoretical framework. In particular, this difference cannot be justified with the aid of the Bell type inequalities (see \cite{BELL} and section
\ref{NLB}).  To distinguish intra-intersystem entanglements, we have to go beyond  quantum theory (see  \cite{PCT1,PQ} and  Appendix 1). 
      
\section{Grangier  et al.  experiment separating classical field theories from quantum mechanics}

We start with citing the  breakthrough paper of Grangier et al.  \cite{GR}:

\footnotesize{ ``However, there has  still  been no test of the conceptually very simple situation dealing with single-photon states of the light impinging on a  beam splitter. In this case, quantum mechanics predicts a  perfect anticorrelation  for photodetections on both  sides of the beam splitter  (a single-photon can only be detected once!),  while any  description involving  classical  fields would predict some amount of coincidences.''} 

Following \cite{GR}, denote by $p_1, p_2$ the probabilities of detection in two channels after beam splitter and by 
 $p_c$ the coincidence probability. Then by using the semiclassical model of detection it is easy to show that  
\begin{equation}
\label{GRE}
p_c \geq p_1 p_2.
\end{equation}
This  inequality  means  clearly that the classical  coincidence probability $p_c$ is always greater than the accidental coincidence probability, which is equal to  $p_1p_2.$ The violation of inequality (\ref{GRE}) thus gives  an  anticorrelation criterion, for characterizing  a nonclassical behaviour of light (see \cite{GR}). 

The crucial theoretical point is that in classical and  semiclassical models the basic physical quantity is intensity of a signal.
In Grangier  et al.  experiment, these are $I(t),$ intensity of imprinting on the beam splitter, and $I_1(t), I_2(t)$ are intensities of signals 
in the two output channels. The use of intensities, instead of counting of clicks, obscures the coincidence detection problem. 
We claim that this is not just a technicality, but the very important foundational issue. And we shall continue discussion in following sections. However, the reader who is not so much interested in foundational questions can jump directly to 
 conclusion-section \ref{CS}.
The main critical point has already been presented.

Finally, we remark PCSFT \cite{PCT1,PQ} suffers of the same problem as  the classical entanglement models - the ``double detection loophole'' (see Appendix 2 for further discussion and  attempts to close this loophole by using the treshold detection scheme). 
  
\section{Quantum measurements}
\label{QMM}

Consider a quantum observable $A$ represented by Hermitian operator $\hat A$ 
with purely discrete spectrum $(a_i).$ By the spectral postulate of quantum mechanics 
any measurement of $A$ produces one of the values $a_i$ (as the result of interaction of 
a quantum system with an apparatus used for $A$-measurement). Thus quantum measurements are characterized 
by individuality of outputs. This crucial feature of quantum measurements was emphasized by Bohr who invented the notion of 
{\it phenomenon} \cite{BR0} (see also \cite{PL1,PL2}): 
 
\footnotesize{`` ... in actual experiments, all observations are expressed by
unambiguous statements referring, for instance, to the registration of the point at which
an electron arrives at a photographic plate. ...  the appropriate physical interpretation of the symbolic quantum mechanical
formalism amounts only to predictions, of determinate or statistical character,
pertaining to individual phenomena ... .} [2,v. 2, p. 64]

Thus, although quantum theory produces statistical predictions, its observables generate individual 
phenomena.  Discreteness of detection events is the fundamental feature of quantum physics justifying existence of quantum systems, carriers of quanta. It is commonly accepted that axiomatic of  quantum theory does not contain the special postulate on discrete clicks and the statistical interpretation of probabilities.\footnote{However, as was pointed by Plotnitsky (unpublished paper),  ``it depends on what he sees as axiomatic of  quantum theory. The structure of complex Hilbert space does not. But once one introduces projectors and the Born's rule to a relate quantum state to the outcome of experiment, both discreteness and probability enter. The Born rule is not part of the Hilbert space structure and, while mathematically natural (in connections the complex quantities of the formalism to real one and the probability) it is brought ad hoc, and why it works, and it works perfectly, is enigmatic. In a way — that is the main quantum mystery — why Born's rule works.''}

One may point to the existence of quantum observables with continuous spectra. The problem of their measurement was analyzed in detail 
by von Neumann \cite{VN}. His analysis implies that measurement of an observable with continuous spectrum has to be reduced to measurements 
of observables with discrete spectra approximating it. This is the complex foundational issue and we  would not go into a deeper discussion;
our considerations are restricted to observables with discrete spectra.

In the classical wave framework the origin of the analog of the quantum Born's rule, so to say the Born's rule for intensities is straightforward. If a classical wave has two orthogonal components, i.e., 
\begin{equation}
\label{I}
\Phi(x)= \Phi_1(x) + \Phi_2(x),
\end{equation}
 with intensities 
$I_1$ and $I_2,$ then corresponding probabilities can be expressed in the form $p_j=I_j/(I_1+I_2), j=1,2,$ and  
intensities are given by the ``classical Born's rule'':
\begin{equation}
\label{I1}
I_j = \Vert \Phi_j\Vert_{L_2}= \int  \vert \Phi_j(x) \vert^2 dx.
\end{equation}
However, this is the separate question whether the coefficients $p_j$ can really be interpreted as probabilities of discrete events. 

In papers on classical entanglement, there are considered expansions of state-vectors with respect to 
orthonormal bases in complex Hilbert spaces.\footnote{The tensor product structure is typically emphasized. 
But this is not the main issue, see section \ref{Le} on classical vs. quantum interpretation of states' superposition.} Such expansions may make the impression that the standard quantum mechanical scheme of measurement can be applied. This is not the case. For classical signals, it is impossible to project the initial state on the state corresponding to one concrete outcome. In the two slit experiment with classical waves, a wave propagating from the source passes both slits at the same time.

Finally, we remark that in classical field theory the method of complex Hilbert space started to be used even before appearance of quantum mechanics. We can mention, for example, {\it Riemann-Silberstein representation}, $\Psi(x)= E(x) + i H(x),$ for the
classical electromagnetic field. In this representation, the Maxwell equation has the form
 of the Schr\"odinger equation. So, studies on classical entanglement are consistent  with complex Hilbert space analysis of classical signals.

\section{Reality without realism (RWR) interpretation  of quantum mechanics}
\label{RWR}

The above discussion on quantum discreteness matches perfectly the RWR interpretation of quantum mechanics that was elaborated in a series 
of works of Plotnitsky (see \cite{SPD} and references herein). This is one of the versions of the Copenhagen interpretation\footnote{Bohr 
had never formulated the Copenhagen interpretation exactly. The quantum community  uses a variety of interpretations pretending to express Bohr's views. 
Therefore, Plotnitsky proposed to speak about interpretations in the spirit of Copenhagen. RWR is one of such interpretations.}. I now present RWR. (This is my interpretation of  RWR. It may differ from Plotnitsky's own views.)

We start by remarking that often Bohr's views are presented as idealism. But, this is  misunderstanding. He definitely 
did not deny reality of {\it quantum  systems}, say electrons or atoms. However, as pointed out in section \ref{QMM}, quantum mechanics does not describe genuine physical properties of quantum systems. Bohr stressed that measurement's 
output cannot be disassociated from a measurement apparatus and generally the complex of experimental physical conditions, experimental context. We can point to two common missuses of quantum theory (well, from the RWR-viewpoint). On one hand, one may neglect the role 
of experimental context and try to assign measurement's output directly to a quantum system. We call this approach ``naive realism''. 
From the Bohr-Heinseberg viewpoint, this approach should be rejected as contradicting Heinseberg's uncertainty relation and generally the complementarity principle. Another  misinterpretation is forgetting about the existence of quantum systems (the reality counterpart of RWR). By Bohr, electron exists! And the output of measurement is assigned to this concrete electron (prepared for measurement), but, of course, the WR-counterpart of Bohr-Plotnitsky interpretation has also to be taken into account. Therefore quantum theory is about such individual assigning of outputs (quantum phenomena).  This is the origin of discreteness of quantum measurements. That is why measurement of intensity of á beam of classical light is not a quantum phenomenon. As was found by realization of the classical entanglement project, generally the WR-counterpart of RWR should be taken into account even for  
classical light. But, surprisingly, the R-counterpart cannot be applied. Hence, classical optics measurements do not produce quantum phenomena in Bohr's meaning.   

Finally, we remark that measurements for a quantum system in the intra-entangled  state satisfies both the R- and WR-counterparts of 
of WRW; so, their outputs are quantum phenomena.

\section{Comparing classical and quantum superpositions}
\label{Le}

From my viewpoint, the misleading journey towards classical entanglement starts already with identifying classical and quantum superpositions. Physically these superpositions are totally different, in spite of the 
possibility to represent them by the same mathematical expression. 

Consider a classical  electromagnetic field with $n$ orthogonal modes corresponding to frequencies $(\nu_j, j=1,...,n)$ with complex amplitudes $(C_j).$ This field can be represented in $n$-dimensional complex Hilbert space with the basis $(e_j\equiv \vert \nu_1\rangle, j=1,...,n):$ 
\begin{equation}
\label{SUP}
\Psi = \sum_j C_j \vert \nu_j\rangle. 
\end{equation}
This vector can be normalized:  
\begin{equation}
\label{SUP1}
\vert \psi\rangle =  \sum_j c_j \vert \nu_j\rangle,
\end{equation}
where $c_j= C_j/\sqrt{ \sum_j \vert C_j\vert^2}.$

\medskip

{\it What is the main difference of classical field superposition (\ref{SUP1}) from the genuine quantum superposition?}

\medskip

The main difference is in measurement procedures determining probabilities $p_j= \vert c_j\vert^2.$ For the classical field, it is impossible to detect discrete clicks in $n$ channels without coincidence detections, where the degree of coincidence is determined by coefficient $g^{(2)}(0).$ Thus, to see the difference between classical and quantum light, on need not consider formal entanglement expressions corresponding for different degrees of freedom. It is sufficient to consider one degree of freedom and superposition. 

\section{Comparing classical and quantum entanglements}
\label{COMP}

Following papers on classical entanglement,  consider two degrees of freedom of the classical electromagnetic field which can be  jointly measured. The four dimensional complex Hilbert space contains states of  this field
that are nonseparable and formally they can be treated as entangled. Here ``entangled'' is understood purely mathematically,
as the special form of representation in complex Hilbert space  endowed with the tensor product structure.
As in the case of superposition, the devil is not in the state, but in measurement. 
For the classical electromagnetic field, photo-detectors cannot produce phenomena, in Bohr's sense. The measurement 
procedure suffers from coincidence detection.   

\section{Quantum information: role of discrete clicks of detectors}

The bit is a portmanteau of binary digit and this preassumes the discrete structure of information represented by bits. 
In Wikipedia, it is stated that  \footnotesize{``a qubit is the quantum version of the classical binary bit that is physically realized with a two-state device. A qubit is a two-state (or two-level) quantum-mechanical system, one of the simplest quantum systems displaying the peculiarity of quantum mechanics.''} Although Wikipedia is not the best source of citation in a research article, this definition is really 
useful for our further analysis. Its second part  reflects common neglect by the role of measurement. At the same time the first part points perfectly to a two-state device, the source of discrete counts; a device that can distinguish two states. Unfortunately, in quantum 
information research one typically operates with states forgetting about extracting information from them. As we have seen in section 
\ref{COMP}, two states superpositions by their selves  are not quantum. Genuine quantum superposition is combination of a state and measurement procedure extracting discrete alternatives.

Thus, quantum information theory is not reduced to linear algebra in complex Hilbert space. Its main component is quantum measurement procedure. The main value of quantum information (as well as classical one) is in the possibility to extract from states discrete 
events.  

\section{Has classical entanglement anything to do with original Bell argument?}

The above critique of attempts to couple studies on classical entanglement with quantum theory can also 
be applied to attempts to couple classical random field correlations violating Bell type inequalities 
with the original Bell argument \cite{B1,B2,B3}. Bell applied classical probability theory to derive his inequality. The 
later was used to compare the classical probabilistic representation of correlations with the quantum theoretical description.

 We recall that in classical probability theory observables are represented 
by random variables, functions on sample space. Denote the latter by symbol  $\Lambda$ (although mathematicians 
typically use symbol  $\Omega).$ Then a random variable $\xi: \Lambda \to  \mathbf{R}$ and by definition of a function it takes only one value $\xi(\lambda)$ for each $\lambda \in \Lambda.$ Thus by getting the clicks 
in both channels one understand that it is impossible to represent such measurements by classical random variables. 

We remark that originally (following EPR-paper \cite{EPR}) Bell was interested in explanation of perfect correlations. In his original inequality \cite{B1}, it was assumed that ranges of values of quantum and classical observables should coincide, i.e., the range of values is the two point set $\{\pm 1\}.$ A classical random variable is a function $\xi: \Lambda \to \{\pm 1\}.$ And if one would accept that for some set of 
$\lambda$s, $\xi$ is multivalued, i.e., at the same time $\xi(\lambda)= -1$ and $\xi(\lambda)= +1,$  then classical probability theory stops to work. There is no way to derive the Bell inequality. In the CHSH-framework, the range of values of observables was extended to the segment $[1, +1].$ However, this was done with only one purpose, namely, to include  value 0 corresponding to non-detection event. Thus in real physical modeling the range of values is given by the discrete set $\{-1, 0, +1\}.$ 

Moreover, as was already pointed out,    von Neumann emphasized \cite{VN} that any Hermitian operator $\hat A$  with continuous spectrum is just a symbolic expression of converging sequence of quantum observables with discrete spectra, representing approximate measurements.

The above remarks on discreteness of quantum and classical observables  were presented only to underline the astonishing difference between measurement procedures in the Bell framework and in classical optics. Even classical random variables with continuous range of values are mathematically represented by single-valued functions.            

\subsection{``Superstrong quantum correlations'': comparing original Bell inequality and CHSH-inequality}

Excitement of researchers violating the CHSH inequality (theoretically or experimentally) with classical field correlations is well understandable.
The statement on ``superstrong quantum correlations'' that cannot be represented as classical correlations has been emphasized in the quantum community.  Typically correlations were associated with states and the issue of quantum vs. classical measurement procedures was practically ignored. 

This is the good place to point that transition from the original Bell inequality \cite{B1} to the CHSH-inequality \cite{CHSH} was not so innocent from the foundational viewpoint. The original Bell inequality is about explicit correlations and hence comparison of 
the concrete values of observables (cf. \cite{EPR}). It is evident that, for this inequality, transition from 
discrete clicks to intensities is nonsense. In the CHSH-framework, this basic issue was obscured. Instead, 
the issue of  ``superstrong quantum correlations'' was elevated (see \cite{AB} for discussion; see also \cite{AE} for related theoretical study).  Nowadays we are much closer to performance of experiments on violation of the original Bell inequality (see \cite{AB} for analysis of the present situation in theory and experiment). Such experiments will immediately distance 
quantum physics from its classical simulation.  

\section{Bell's inequalities as tests of  observables' incompatibility}
\label{NLB}

The unconventional interpretation of Bell's type inequalities was proposed in recent author's paper \cite{BELL}. 
This paper presents purely quantum mechanical treatment of these inequalities, i.e., without 
any relation to hidden variables. Observables measured in experiments are coupled directly to 
quantum observables.  It was shown that in this framework these {\it  inequalities express the compatibility-incompatibility 
interplay for local observables.} Thus quantum theory has nothing to do with nonlocality. For reader's 
convenience we briefly present the aforementioned analysis.     

The  quantum theoretical CHSH-correlation function has the form:
\begin{equation}
\label{LC}
\langle  {\cal B} \rangle_{\psi}  =\frac{1}{2} [\langle \hat A_1 \hat B_1  \rangle_{\psi} + \langle \hat A_1 \hat B_2 \rangle_{\psi} + \langle \hat A_2  \hat B_1  \rangle_{\psi} - \langle \hat A_2 \hat B_2 \rangle_{\psi}],
\end{equation}
where $\psi$ is a pure quantum state (mixed states can be considered as well). 
(This quantum theoretical correlation functions  is compared with the experimental CHSH-correlation function.)

In the quantum framework,  the CHSH-correlation function can be expressed with the aid of the Bell-operator:
\begin{equation}
\label{L1}
\hat {\cal B} = \frac{1}{2}[\hat A_1(\hat B_1+ \hat B_2) +\hat A_2(\hat B_1-\hat B_2)]
\end{equation}
as
\begin{equation}
\label{L1T}
\langle  {\cal B} \rangle_{\psi}= \langle \psi\vert \hat {\cal B} \vert \psi\rangle. 
\end{equation}

By straightforward calculation one can derive the Landau identity: 
\begin{equation}
\label{L2}
\hat{{\cal B}}^2=I - (1/4) [\hat A_1, \hat A_2][\hat B_1,\hat B_2].
\end{equation}

This  identity implies that if at least one of commutators $[\hat A_1, \hat A_2], [\hat B_1,\hat B_2]$ 
equals zero, i.e., if at least one pair of observables, $(A_1, A_2)$ or (and) $(B_1, B_2),$ is compatible, 
then for any state $\psi,$  
\begin{equation}
\label{L1T}
\sup_{\Vert \psi \Vert=1} \vert \langle  {\cal B} \rangle_{\psi}\vert= \Vert \hat{{\cal B}} \Vert =1, 
\end{equation}
i.e., for each state $\psi,$
\begin{equation}
\label{L1Ta}
 \vert \langle  {\cal B} \rangle_{\psi}\vert= \Vert \hat{{\cal B}} \Vert \leq 1.
\end{equation}
This is the quantum version of the CHSH-inequality. The classical bound by 1 has the purely  quantum explanation.

Simple spectral analysis shows (see \cite{}) that if the product of commutators is not equal to 
zero, i.e., in both pairs  $(A_1, A_2)$ and $(B_1, B_2)$ of observables are incompatible, 
then either
\begin{equation}
\label{L1Tzm}
\sup_{\Vert \psi \Vert=1} \vert \langle  {\cal B} \rangle_{\psi}\vert= \Vert \hat{{\cal B}} \Vert >1, 
\end{equation}
or, for  $\hat{{\cal B}}_-= \frac{1}{2}[\hat A_1(\hat B_2 - \hat B_1) +\hat A_2(\hat B_1-\hat B_2)],$ 
\begin{equation}
\label{L1Tzmk}
\sup_{\Vert \psi \Vert=1} \vert \langle  {\cal B}_- \rangle_{\psi}\vert= \Vert \hat{{\cal B}}_- \Vert >1 
\end{equation}

This condition can be rewritten in a compact form. Denote by $\sigma$ some permutation of the indexes 
for the $A$-observables and the indexes for the $B$-observables and denote by $\hat{{\cal B}}_\sigma$ the operator 
with corresponding permutation of indexes. If the  product of  commutators is not equal to 
zero, then
\begin{equation}
\label{L1Tzmh}
\max_\sigma \sup_{\Vert \psi \Vert=1} \vert \langle  {\cal B}_\sigma \rangle_{\psi}\vert >1, 
\end{equation}
i.e., there exists some state $\psi$ such that the CHSH-inequality  is violated at least for one 
of correlations $\langle  {\cal B}_\sigma \rangle_{\psi}.$

The issue of locality can be formalized by introducing the tensor product structure on the state space $H,$
i.e., $H=H_1\otimes H_2$ and  considering observables represented by Hermitian operators in the form
$\hat A_i = \hat {\bf A}_i \otimes I$ and $\hat B_i = I  \otimes \hat {\bf B}_i,$ where Hermitian operators 
$\hat {\bf A}_i$ and $\hat {\bf B}_i$ act in spaces $H_1$ and $H_2,$ respectively. Then the condition of commutativity respects the tensor product structure, since $[A_1, A_2]= [\hat {\bf A}_1, \hat {\bf A}_2]\otimes I $ and  $[B_1, B_2]= I \otimes [\hat {\bf B}_1, \hat {\bf B}_2].$

Now, if the tensor product structure corresponds to the compound system structure, then 
$[\hat {\bf A}_1, \hat {\bf A}_2]\not=0$ and $[\hat {\bf B}_1, \hat {\bf B}_2]\not=0$ are conditions of {\it local
incompatibility of observables.} Thus satisfaction-violation of the CHSH-inequality is completely determined 
by these local conditions.\footnote{In the absence of the tensor product structure, we have to impose the constraint that
the product of commutators differs from zero. The presence of tensor product structure makes this constraint redundant.}

{\it By interpreting the Bell type inequalities as describing the compatibility-incompatibility interplay we cannot point to any difference between ``intersystem and intrasystem entanglement''.}

At the same time, analysis presented in the previous sections 
points to crucial difference between classical and quantum entanglements.\footnote{This is the reply to the question of Mario Krenn
who told me about extended studies on classical entanglement and asked me to comment them in the light 
the purely quantum mechanical analysis of the Bell type inequalities (after my talk at FQMT19, Prague).} For classical light, the presented incompatibility interpretation of the CHSH inequality for quantum systems has to used with caution. We restrict considerations to intra-entanglement. Of course, the mathematical structure of states and operators is the same. Thus, all above calculations are valid even 
in the classical entanglement framework. However, the physical meaning of operators is not the same. In quantum physics, the operators represent measurements procedures which do not suffer of the double detection loophole; in classical optics, the same operators
represent measurements procedures which suffer of this loophole. It is not clear whether one can extend the complementarity principle 
to such measurements. (This is the good question to experts in quantum foundations). 

PCSFT reproduces quantum correlations without establishing isomorphism of state spaces and physical variables, subquantum$\to$quantum 
map has a more complex structure. Therefore the above operator analysis of the CHSH-inequality has no direct impact to this theory.
To couple consequences of this analysis with PCSFT, we have to understand how complementarity arises through transition from a subquantum 
theory to quantum mechanics.

\section{Concluding remarks}
\label{CS}

The aim of this note is to distance the technical impact of ``classical entanglement'' research (both for theory and experiment) \cite{REV}
from its misleading interpretation, as supporting ``quantum nonlocality'' \cite{NL}. 
First we present the main points of our analysis of the notion ``quantum nonlocality'':
\begin{itemize}
\item In modern physics, its using is the real mess.
\item The ontic-epistemic (descriptive-observational) viewpoint on scientific theories clarifies misunderstanding 
between Einstein and Bohr.
\item Einstein's treatment of elements of reality as components of observational theory leads him to really misleading 
notion of quantum nonlocality, based on a spooky action at a distance.
\item Bell type inequalities have the purely quantum interpretation as tests of local incompatibility.
\end{itemize}

We now list the main conclusions from our analysis of interrelation of classical and  quantum entanglements: 
\begin{itemize}
\item The  main issue is the difference between classical and quantum superpositions. 
It can be explained by Grangier  et al.  experiment  \cite{GR}
\item The distinguishing feature of quantum measurements is discreteness and 
individuality of outcomes (as expressed in Bohr's notion of phenomenon).  
  \item  Derivation of quantum(-like) correlations with classical entanglement \cite{NL,REV} implies 
that  the Hilbert space formalism has to be distinguished from genuine quantum physics.
\item Classical entanglement is not consistent with Bell's hidden variables theory:
 coincidence detection blocks the use of  random variables.
\end{itemize} 
This comparison of classical and quantum entanglements and critique of the ``quantum nonlocality'' interpretation 
of their difference is the main output of the paper.

Finally, we point to the recent review of  Korolkova and Leuchs \cite{GL} which is the important step towards resolution of the interpretation problems related to 
interrelation of classical and quantum entanglements. Its authors do not more refer to quantum nonlocality (cf. with previous review \cite{REV}). They recognize that the main issue is not the impossibility to generate intersystem entanglement with classical optics. The main issue is the difference between intra system entanglements, classical vs. quantum. And   Korolkova and Leuchs, as well as the author of this paper, also emphasize the role of measurement procedures in distinguishing two types of entanglement. They made the following remark on 
an intra-entangled state of a genuine quantum system: {\it ``This state is the quantum entangled state of type
$|01>+|10>$ a strict correlation of one photon in one arm and no photon in the other or vice versa.''} I would just add that, 
in fact, the root of the problem lies already in classical vs. quantum interpretation of superposition-state $|0>+|1>.$

\section*{Appendix 1: Subquantum modeling of inter-intra system entanglements}
		
One possibility is to appeal to  so-called {\it prequantum classical random field theory} (PCSFT) \cite{PCT1,PQ} that is devoted to modeling of {\it both forms of entanglement}, intra and inter system,  with the aid of  classical random fields.  PCSFT provides the abstract random field representation of quantum averages and correlations. In PCSFT,  intra and inter system entanglements have different mathematical representations. The crucial point is that representation of intersystem entanglement (in PCSFT) is impossible without assuming the presence of {\it a random background field}, a kind of the zero point field (field of vacuum fluctuations) explored in stochastic electrodynamics. In principle, the presence of such a background field can be interpreted as nonlocality, although the use of such a terminology would be really misleading. Say in radiophysics, nobody would associate some mystical features with a random  background. However, in this note we shall not present the details of the PCSFT modeling of intra and inter system entanglements. We plan to do this in a future publication.

\section*{Appendix 2: Extracting discrete events from continuous random fields}

We remark that the Grangier-type experiments were directed against one special 
model of photo-detection, the semiclassical model (see \cite{MANDEL}). ``In the semiclassical theory of 
photoelectric detection, it is
found that the conversion of continuous electromagnetic radiation
into discrete photoelectrons is a random process.'' (see \cite{GRT}) One can propose other detection
models for such conversion. The simplest way to extract discrete events from continuous random fields is to use  threshold detection procedures. Such a project was started in the PCSFT-framework \cite{PCT1,PQ,PCT6,PCT5}. 
First we consider intrasystem entanglement. In this case, the threshold detection scheme
can be designed to exclude the double detection (clicks in both channels) for a dichotomous observable. Here theoretical research was completed by numerical simulation  \cite{PCT6}. The coefficient of second order coherence $g^{(2)}(0)$ is used as the measure ``quantumness''; it is possible to construct such classical random fields and the threshold detection scheme that $g^{(2)}(0)< 1.$ 

Now consider intersystem entanglement, in the PCSFT-realization. This realization can also be equipped with a threshold detection scheme and 
correlations based on probabilities for discrete counts can violate the CHSH-inequality \cite{PCT5}. However, in this case I was not able to 
close the double detection loophole. The model of (classical field based)  intersystem entanglement endowed with the threshold detection scheme is so complex \cite{PCT5} that it is difficult to estimate the magnitude of coefficient $g^{(2)}(0).$

\section*{Acknowledgments} 

This work was supported by the research project of the Faculty of Technology, Linnaeus University, 
``Modeling of complex hierarchical structures''.

 \end{document}